\documentclass[usenatbib]{mn2e}
\usepackage{epsfig}
\usepackage{amsmath,amssymb}
\usepackage{aas_macros}
\newcommand{\be}{\begin{equation}}
\newcommand{\ee}{\end{equation}}
\newcommand{\bea}{\begin{eqnarray}}
\newcommand{\eea}{\end{eqnarray}}
\newcommand{\hMsol}{{\,h^{-1}\rm M}_\odot}
\newcommand{\hkpc}{{\,h^{-1}\rm kpc}}
\newcommand{\kms}{{\rm km/s}}
\title{Supersonic Motions of Galaxies in Clusters}
\author[Faltenbacher et al.]
{Andreas Faltenbacher$^{1,3}$, Andrey V. Kravtsov$^2$, Daisuke Nagai$^2$, Stefan Gottl\"ober$^3$
\\
\\
$^1$Racah Institute of Physics, Hebrew University, Jerusalem 91904, Israel 
\\
$^2$Department of Astronomy and Astrophysics,
       Kavli Institute for Cosmological Physics,
       The University of Chicago, Chicago, IL 60637
\\
$^3$Astrophysikalisches Institut Potsdam, 
An der Sternwarte 16, 14482 Potsdam, Germany
}
\date{\today}
\begin{document}
\maketitle
\begin{abstract}
  We study motions of galaxies in galaxy clusters formed in the
  concordance $\Lambda$CDM cosmology. We use high-resolution
  cosmological simulations that follow dynamics of dark matter and gas
  and include various physical processes critical for galaxy
  formation: gas cooling, heating and star formation. Analysing
  motions of galaxies and the properties of intracluster gas in the
  sample of eight simulated clusters at $z=0$, we study velocity
  dispersion profiles of the dark matter, gas, and galaxies. We
  measure the mean velocity of galaxy motions and gas sound speed as a
  function of radius and calculate the average Mach number of galaxy
  motions. The simulations show that galaxies, on average, move
  supersonically with the average Mach number of $\approx 1.4$,
  approximately independent of the cluster-centric radius.  The
  supersonic motions of galaxies may potentially provide an important
  source of heating for the intracluster gas by driving weak shocks
  and via dynamical friction, although these heating processes appear
  to be inefficient in our simulations. We also find that 
  galaxies move slightly faster than the dark matter particles.  The
  magnitude of the velocity bias, $b_v\approx 1.1$, is, however,
  smaller than the bias estimated for subhalos in dissipationless
  simulations.  Interestingly, we find velocity bias in the tangential
  component of the velocity dispersion, but not in the radial
  component. Finally, we find significant random bulk motions of
  gas. The typical gas velocities are of order $\approx 20-30\%$ of
  the gas sound speed.  These random motions provide about $10\%$ of
  the total pressure support in our simulated clusters. The
  non-thermal pressure support, if neglected, will bias measurements
  of the total mass in the hydrostatic analyses of the X-ray cluster
  observations.
\end{abstract}      

\section{Introduction}

Clusters of galaxies are powerful probes of both the overall cosmic
evolution and evolution of galaxies in dense environments.  Most of
the cluster baryons are in the form of hot diffuse plasma and stars in
galaxies, with a fraction of $\sim 10-50\%$ of stars in the
intracluster medium. Interaction between galaxies and the hot
intracluster gas is thought to be important in shaping properties of
both components.  The compression of the interstellar medium (ISM) of
an accreted galaxy and cluster tidal forces can lead to a
starformation burst
\citep{dressler_gunn83,gavazzi_etal95,rose_etal01,koopmann_kenney04,sakai_etal02}
or active galactic nuclei (AGN) activity
\citep[e.g.,][]{miller_owen03} and morphological transformation
\citep{moore_etal96,koopmann_kenney04,gnedin03b}. The subsequent
removal of the ISM by ram pressure stripping \citep{gunn_gott72} and
other processes \citep[e.g.,][]{nulsen82} can suppress starformation,
dramatically affecting properties of cluster galaxy population
\citep[e.g.,][]{quilis_etal00,schulz_struck01,vollmer_etal01,vollmer03}.
Galaxies, in turn, may heat the intracluster medium (ICM) via
supernova or AGN feedback and enrich it with heavy elements
\citep[e.g.,][]{kaiser91,metzler_evrard94,valageas_silk99,churazov_etal02,fabian_etal02,scannapieco_oh04,voit_ponman03,ruszkowski_etal04}.
Galaxies can stir the surrounding gas as they move through the ICM
driving turbulent flows and mixing the gas.  The kinetic energy of
galaxy motions can also be an important source of the ICM heating
\citep*[e.g.,][]{miller86,elzant_etal04}.

In the currently favoured Cold Dark Matter (CDM) model of structure
formation, galaxies form in the extended dark matter halos as gas
cools and condenses in the centre. After a galaxy is accreted by a
cluster, the tidal forces can quickly strip the outer loosely bound
regions of DM halos and can lead to a total disruption of the halo
and, possibly, the stellar system. The dynamical evolution of
galaxy-size dark matter halos in groups and clusters (often called
{\it the subhalos}) has been the subject of many recent studies, which
used a new generation of high-resolution dissipationless simulations
not affected by the ``overmerging'' problem
\citep{ghigna_etal98,tormen_etal98,klypin_etal99,colin_etal99,okamoto_habe99,
colin_etal00,ghigna_etal00,delucia_etal04,kravtsov_etal04,desai_etal04,diemand_etal04,gao_etal04a,reed_etal04}.

These studies find that abundance of subhalos is in reasonable
agreement with the observed abundance of cluster galaxies
\citep[][see, however, \citeauthor{diemand_etal04}
\citeyear{diemand_etal04} and \citeauthor{gao_etal04b}
\citeyear{gao_etal04b}]{moore_etal99,kravtsov_etal04,desai_etal04},
although there are suggestive differences in the radial distribution
\citep[][]{diemand_etal04,gao_etal04b,nagai_kravtsov04} and circular
velocity functions \citep{desai_etal04}. Interestingly, velocity
dispersion of subhalos is larger than that of dark matter by a factor
up to $\approx 1.4$ in the inner regions of clusters
\citep{colin_etal00,ghigna_etal00,diemand_etal04,gao_etal04b},
implying that dynamical estimates of cluster masses using galaxy
velocity dispersion may be biased.  However, applying results of
dissipationless simulations to the properties of cluster galaxies is
subject to many caveats. Gas cooling during galaxy formation increases
central density of DM halos
\citep[e.g.,][]{blumenthal_etal86,gnedin_etal04}, making them more
resistant to tidal disruption. In the inner regions of galaxies
stellar density is typically considerably higher than that of DM and
the luminous component can survive the tidal forces even if the
surrounding DM halo is completely disrupted
\citep[e.g.,][]{gao_etal04b}. It is therefore critical to study the
properties of cluster galaxies forming in self-consistent cosmological
simulations.

Although a number of studies during the last decade used gasdynamics
cosmological simulations with cooling to study spatial distribution of
galaxies \citep[e.g.,][]{katz_etal92,pearce_etal99,
  yoshikawa_etal01,weinberg_etal04,zheng_etal04}, only a few analyses directly
addressed the motions of galaxies in cluster halos.
\cite{frenk_etal96} used simulations with cooling and starformation to
study distribution and dynamics of galaxies in clusters. They found
that the velocity dispersion of massive galaxies in the simulations is
lower than that of the DM by $\sim 20-30\%$, which they attributed to
the orbital energy loss due to dynamical friction.  More recently,
\cite{berlind_etal03} analysed mean motions of galaxies,
identified as dense baryonic clumps, in halos of different masses in SPH
simulations. These authors find that galaxies selected to have
baryonic masses above a certain threshold move slower than DM in galaxy
and group-size halos, the bias that disappears for more massive 
cluster-size halos. 
 
In this paper we present analysis of a sample of eight galaxy clusters
formed in high-resolution cosmological simulations of the $\Lambda$CDM
model. The simulations follow dynamics of DM and gas and include
relevant cooling and heating processes and star formation. 
We use a combined sample of cluster galaxies, constructed using 
the friends-of-friends (FoF) algorithm, to study the average statistical 
properties of galaxy motions and compare them to the corresponding
statistics calculated for the DM and intracluster gas. A complementary
study of the radial distribution of galaxies in these simulations is presented
in \citet{nagai_kravtsov04}. 

The paper is organised as follows. In \S~\ref{sec:sim} we introduce
the simulation and describe the cluster sample which the subsequent
analysis is based on. In \S~\ref{sec:galaxies} we describe details of
the galaxy finding algorithm and present analysis
of the overall spatial and velocity distributions of galaxies. In
\S~\ref{sec:motions} we study the radial profiles of the velocity
dispersion and velocity anisotropy for the DM, gas, and galaxies.
The radial dependence of the average Mach number of
galaxy motions in simulated clusters is presented in
\S~\ref{sec:mach}. Finally, in \S~\ref{sec:discussion} we summarise
our main results and conclusions.

\section{Simulated cluster sample }
\label{sec:sim}
\subsection{Simulations}

In this study, we analyse high-resolution cosmological simulations of
eight group and cluster-size systems formed in the ``concordance''
flat {$\Lambda$}CDM model: $\Omega_{\rm m}=1-\Omega_{\Lambda}=0.3$,
$\Omega_{\rm b}=0.021h^{-2}$, $h=0.7$ and $\sigma_8=0.9$, where the
Hubble constant is defined as $100h{\ \rm km\ s^{-1}\ Mpc^{-1}}$, and
$\sigma_8$ is the power spectrum normalisation on $8h^{-1}$~Mpc
scale. The simulations were done with the Adaptive Refinement Tree
(ART) $N$-body$+$gasdynamics code \citep*{kravtsov_etal97,kravtsov99,
kravtsov_etal02}, a Eulerian code that uses the adaptive refinement
in space and time and (non-adaptive) refinement in mass
\citep{klypin_etal01} to reach the high dynamic range required to
resolve cores of halos formed in self-consistent cosmological
simulations.

To set up initial conditions we first ran a low resolution simulation
of an $80h^{-1}$~Mpc box and selected eight clusters.  The virial
masses of clusters we selected range from $\approx
7\times10^{13}$ to $3\times 10^{14}h^{-1}{\ \rm M_{\odot}}$.  The
perturbation modes in the lagrangian region corresponding to the
sphere of five virial radii around each cluster at $z=0$ have then been
re-sampled at the initial redshift, $z_i=49$, retaining the previous
large-scale waves intact but including additional small-scale waves,
as described by \citet{klypin_etal01}.

High-resolution simulations were run using 128$^3$ uniform grid and 8
levels of mesh refinement in the computational box of $80h^{-1}$~Mpc,
which corresponds to the dynamic range of $128\times 2^8=32768$ and
peak formal resolution of $80/32,768\approx 2.44h^{-1}\ \rm kpc$,
corresponding to the actual resolution of $\approx 2\times 2.44\approx
5h^{-1}\ \rm kpc$. Only the region of $\sim 10h^{-1}\ \rm Mpc$ around
the cluster was adaptively refined, the rest of the volume was
followed on the uniform $128^3$ grid. The dark matter particle mass in
the region around the cluster was $2.7\times 10^{8}h^{-1}{\rm\ 
  M_{\odot}}$, while other regions were simulated with lower mass
resolution.

As the zeroth-level fixed grid consisted of only $128^3$ cells, we
started the simulation already pre-refined to the 2nd level
($l=0,1,2$) in the high-resolution lagrangian regions of clusters.
This is done to ensure that the cell size is equal to the mean
interparticle separation and all fluctuations present in the initial
conditions are evolved properly. During the simulation, the
refinements were allowed to the maximum $l=8$ level and refinement
criteria were based on the local mass of DM and gas in each cell. The
logic is to keep the mass per cell approximately constant so that the
refinements are introduced to follow the collapse of matter in a
quasi-lagrangian fashion. For the DM, we refine the cell if it
contains more than two dark matter particles of the highest mass
resolution specie.  For gas, we allow the mesh refinement, if the cell
contains gas mass larger than four times the DM particle mass scaled
by the baryon fraction.  In other words, the mesh is refined if the
cell contains the DM mass larger than $5.42\times 10^8h^{-1}\ {\rm
M_{\odot}}$ $=2(1-f_b)m_p$ or the gas mass larger than $1.81\times
10^8h^{-1}\ {\rm M_{\odot}}$ $=4f_bm_p$, where
$m_p=3.16\times10^8h^{-1}\ {\rm M_{\odot}}$ and $f_b = \Omega_{\rm
b}/\Omega_{\rm m}$ = 0.1429.  We analyse clusters at the present-day
epoch as well as their progenitors at higher redshifts. 

Simulations included gasdynamics and several physical processes
critical to various aspects of galaxy formation: star formation, metal
enrichment and thermal feedback due to the supernovae type II and type
Ia, self-consistent advection of metals, metallicity-dependent
radiative cooling and UV heating due to cosmological ionising
background \citep{haardt_madau96}.  The cooling and heating rates take
into account Compton heating/cooling of plasma, UV heating, atomic and
molecular cooling and are tabulated for the temperature range
$10^2<T<10^9$~K and a grid of metallicities, and UV intensities using
the {\tt Cloudy} code \citep[ver. 96b4,][]{ferland_etal98}. The Cloudy
cooling and heating rates take into account metallicity of the gas,
which is calculated self-consistently in the simulation, so that the local
cooling rates depend on the local metallicity of the gas.

Star formation in these simulations was done using the
observationally-motivated recipe \citep[e.g.,][]{kennicutt98}:
$\dot{\rho}_{\ast}=\rho_{\rm gas}^{1.5}/t_{\ast}$, with
$t_{\ast}=4\times 10^9$~yrs. Stars are allowed to form in regions with
temperature $T<2\times10^4$ K and gas density $n > 0.1\ \rm cm^{-3}$.
No other criteria (like the collapse condition $\nabla\cdot {\bf v} <
0$) were used.  Algorithmically, star formation events are assumed to
occur once every global time step $\Delta t_0\sim 10^7$ yrs, the value
close to the observed timescales \citep[e.g.,][]{hartmann02}.
Collisionless stellar particles with mass
$m_{\ast}=\dot{\rho}_{\ast}\Delta t_0$ are formed in every unrefined
mesh cell that satisfies criteria for star formation during star
formation events. The mass of stellar particles is restricted to be
larger than $\min(5\times 10^7h^{-1}{\ \rm M_{\odot}},2/3\times m_{\rm
gas})$, where $m_{\rm gas}$ is gas mass in the star forming cell.
This is done in order to keep the number of stellar particles
computationally tractable, while avoiding sudden dramatic decrease of
the local gas density. In the simulations analysed here, the masses of
stellar particles formed by this algorithm range from $\approx 10^5$
to $7\times 10^8h^{-1}{\ \rm M_{\odot}}$.
  
Once formed, each stellar particle is treated as a single-age stellar
population and its feedback on the surrounding gas is implemented
accordingly.  The feedback here is meant in a broad sense and includes
injection of energy and heavy elements (metals) via stellar winds and
supernovae and secular mass loss.  Specifically, in the simulations
analysed here, we assumed that stellar initial mass function (IMF) is
described by the \citet{miller_scalo79} functional form with stellar
masses in the range $0.1-100\ \rm M_{\odot}$. All stars more massive
than $M_{\ast}>8{\ \rm M_{\odot}}$ deposit $2\times 10^{51}$~ergs of
thermal energy in their parent cell\footnote{No delay of cooling was
introduced in these cells.} and fraction $f_{\rm Z}= {\rm
min}(0.2,0.01M_{\ast}-0.06)$ of their mass as metals, which crudely
approximates the results of \citet{woosley_weaver95}. In addition, the
stellar particles return a fraction of their mass and metals to the
surrounding gas at a secular rate $\dot{m}_{\rm
loss}=m_{\ast}\,\,C_0(t-t_{\rm birth} + T_0)^{-1}$ with $C_0=0.05$ and
$T_0=5$~Myr \citep{jungwiert_etal01}. The code also accounts for SNIa
feedback assuming a rate that slowly increases with time and broadly
peaks at the population age of 1~Gyr. We assume that a fraction of
$1.5\times 10^{-2}$ of mass in stars between 3 and $8\ \rm M_{\odot}$
explodes as SNIa over the entire population history and each SNIa
dumps $2\times 10^{51}$ ergs of thermal energy and ejects $1.3\ \rm
M_{\odot}$ of metals into parent cell. For the assumed IMF, 75 SNII
(instantly) and 11 SNIa (over several billion years) are produced by a
$10^4\ \rm M_{\odot}$ stellar particle.

\subsection{Cluster sample}
\label{sec:cluster}

Table~\ref{tab:today} lists properties of the eight clusters used in
this study.  The mass of each cluster is defined within the radius
enclosing the cumulative density of $500\rho_{crit}$, where
$\rho_{crit}$ is the critical density of the universe. This choice of
overdensity is motivated by the fact that clusters are on average more
relaxed in their inner regions \citep*{evrard_etal96}. Additionally,
most of the X-ray radiation comes from region within $r_{500}$ so this
radius is preferred in X-ray observations. In the following we use the
radius $r_{500}$, mass $M_{500}\equiv M(<r_{500})$, and circular
velocity at $r_{500}$, $v_{500}\equiv\sqrt{GM_{500}/r_{500}}$, to
normalise the physical cluster-centric radii, masses, and velocities.
We will use the normalised quantities to obtain statistics averaged
over all simulated clusters.
\begin{table}
\begin{center}
\begin{tabular}{cccccc}
\hline
id& $r_{500}$& $v_{500}$& $m_{dark}$& $m_{gas}$& $m_{star}$\\\hline\hline
   1& 606& 963& 1.111&  0.105&  0.091\\\hline
   2& 661&1047& 1.431&  0.137&  0.115\\\hline
   3& 621& 988& 1.202&  0.120&  0.085\\\hline
   4& 520& 830& 0.706&  0.074&  0.054\\\hline
   5& 486& 771& 0.570&  0.043&  0.059\\\hline
   6& 535& 851& 0.763&  0.078&  0.061\\\hline
   7& 506& 806& 0.648&  0.061&  0.055\\\hline
   8& 465& 743& 0.509&  0.044&  0.045\\\hline\hline
mean& 550& 875& 0.867&  0.083&  0.070\\\hline
\end{tabular}
\caption{\label{tab:today}
Characteristic properties of the clusters. Column description: (1) the
identification number of each cluster; (2) the radius $r_{500}$
enclosing overdensity of $500\times\rho_{crit}$ in $\hkpc$; (3)
circular velocity at $r_{500}$:
$v_{500}\equiv\sqrt{GM_{500}/r_{500}}$  in $\kms$; (4-6) the mass 
of dark matter, gas, and stars within $r_{500}$
in units of $10^{14}\hMsol$. The last row shows the mean
values averaged over all eight clusters.       
}
\end{center}
\end{table}
\section{Galaxy sample}
\label{sec:galaxies}
\subsection{Galaxy identification}
\label{sec:galid}

We use the friends-of-friends (FoF) algorithm
\citep[e.g.,][]{einasto_etal84} to identify galaxies in the
simulations. The galaxies are defined as groups of stellar particles
linked by the FoF with a certain linking length. We should note that
the use of the FoF algorithm in dense environments can lead to
misidentification of systems in two ways. First, distinct objects
connected by an accidental bridge of small number of particles can be
identified by the algorithm as a single object.  Second, the algorithm
can identify an unbound group of particles which does not correspond
to a real physical system.  To minimise such problems, we adopt a
two-step clustering analysis.

In the first step, we identify groups with the FoF linking length of
$ll = 5\hkpc$. This empirical value is similar to the typical sizes of
galaxies and is well below the average distance between galaxies in a
cluster. It is also close to the peak spatial resolution in the
simulation. In the second step, we consider identified groups of more
than $100$ stellar particles and apply the FoF analysis with a smaller
linking length of $ll=2\hkpc$ only to particles belonging to these
groups. For our analysis, we then consider only identified stellar
particle groups (galaxies) with more than $50$ particles linked with
$ll=2\hkpc$.

\begin{figure}
\epsfig{file=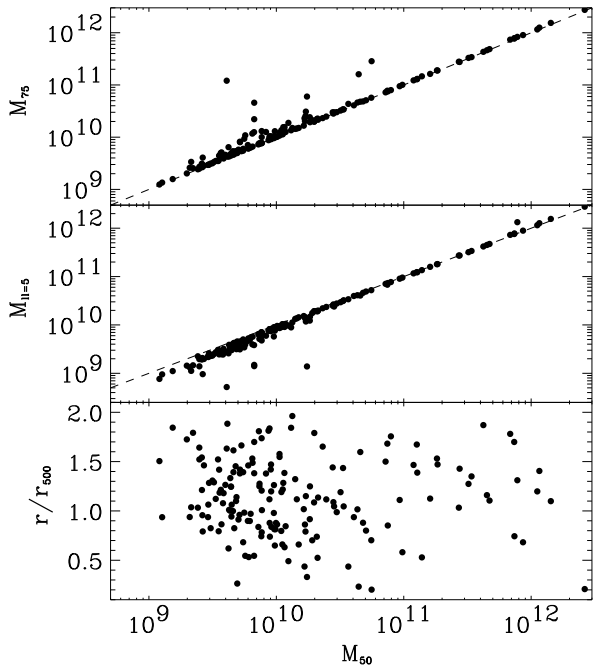,width=0.92\hsize}  
\caption{\label{fig:mass} The data points represent the properties of
all the galaxies found in the eight simulated clusters at z=0 using
the approach described in section~\ref{sec:galid}. The upper panel
compares the stellar mass content within spheres of radius $50\hkpc$
and $75\hkpc$. The data points which coincide with the dashed diagonal
belong to galaxies which contain all of their stellar mass within a
radius of $50\hkpc$. The middle panel shows a comparison between the
masses ${\rm M}_{ll}$ obtained by FOF with $ll = 5 \rm{kpc}/h$ and the
masses within a sphere of radius $50\hkpc$. Points located on the
dashed diagonal represent $100\%$ consistency of the masses obtained by
the FoF approach and the masses gathered within a sphere of
$50\hkpc$. The lower panel depicts the relation between galaxy mass
and distance to the cluster centre. Where the distance of each galaxy
is scaled by $r_{500}$ (see Table~\ref{tab:today}) of the hosting
cluster.}
\end{figure}

Figure~{\ref{fig:mass} compares the stellar masses, $M_{50}$ and $M_{75}$, 
within the spheres of $r_{50}=50\hkpc$ and $r_{75} = 75\hkpc$ radii 
around the identified stellar groups in all eight clusters. The
upper panel of figure~\ref{fig:mass} shows that the dominant fraction
of stellar mass resides within a radius of $50\hkpc$.  In a few cases
the $M_{75}$ substantially exceeds $M_{50}$. These rare cases
correspond to close encounters between galaxies resulting in a
significant overlap of their stars. This occurs most often when a
small galaxy passes near the central cluster galaxy.  The comparison
of the FOF masses $M_{ll}$ and $M_{50}$ shows that the FoF with
$ll=5\hkpc$ correctly links all of the stellar mass in galaxies. The
small deviation from the $M_{ll}=M_{50}$ line at low masses is due to
the contamination by the background particles. The background of
intracluster stars and stars associated with the central galaxies at
small masses provides a small contribution to the mass within
$50\hkpc$ for small sized objects. The contribution is evidently small
however for our purposes and we did not attempt to correct for it.

The lower panel of figure~\ref{fig:mass} shows the cluster-centric
distance of the individual galaxies (scaled by the $r_{500}$ of the
corresponding cluster) versus ${\rm M}_{50}$. There is a small
deficiency of low mass ($M_{50}\lesssim 5\times 10^9\hMsol$) and high
mass ($M_{50}\gtrsim 2\times 10^{11}\hMsol$) galaxies at small
distances. The small mass galaxies are likely affected by the limited
numerical resolution and tidal force of the cluster which leads to the
tidal mass and possibly complete disruption. The minimum particle
number limit we imposed can then exclude such objects.  At large
masses, the deficiency is likely due to the efficient merging via
dynamical friction.

\subsection{Galaxy number density profile}
\label{sec:nrden}

In our analysis below we use the average number density profile of the
compiled galaxy sample. The profile is shown in
figure~\ref{fig:dens.lin010}. For comparison we also show the average
DM and gas density profiles. The profiles are calculated by averaging
profiles of individual clusters with radii normalised to $r_{500}$ and
densities normalised to the mean densities within $r_{500}$.
\begin{figure}
\epsfig{file=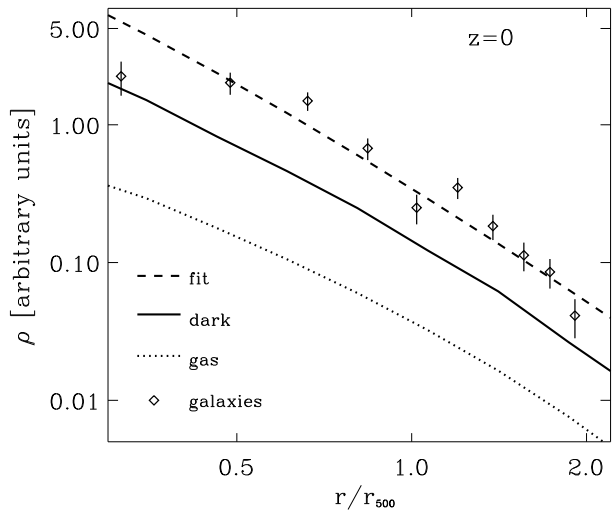,width=0.95\hsize} 
\caption{\label{fig:dens.lin010}
  Average density profiles of galaxy number density ({\it diamond}
  symbols) and density of dark matter ({\it solid} line) and gas ({\it
    dotted} line). The error bars for galaxy number density indicate
  the $1\sigma$ confidence interval. The {\it dashed} line is a 
  fit of the analytic NFW profile (eq.~[\ref{equ:notnfw}]). The relative
normalisation of the DM, gas, and galaxy profiles is arbitrary.}
\end{figure}
We fit the galaxy number density profile with the 
analytic NFW profile \citep{navarro:profI,navarro:profII}:
\be
\label{equ:notnfw}
n_{\rm gal}(r) = {n_0\over(r/r_s)(1+r/r_s)^2}\ .
\ee
The galaxy profile is well described by the NFW profile at
$r/r_{500}\gtrsim 0.5$ but is flatter than the profile of DM at
smaller radii. The difference may be caused by tidal mass loss and
disruption experienced by galaxies in the core of the cluster.  The
difference is rather small, however. The overall galaxy distribution
is rather similar to that of dark matter at the radii we reliably
probe.  The detailed analysis of the galaxy number density profiles in
these simulations and comparisons with observations are presented in
\citet{nagai_kravtsov04}.

\subsection{Velocity distribution}
\label{sec:mabo}

Before we proceed to the analysis of average galaxy 
motions, it is important to consider velocity distribution of
cluster galaxies.  If the velocity distribution follows the
Maxwell-Boltzmann distribution both the velocity dispersion
and the mean velocity are meaningful and are connected by
the well-defined relation. Figure~\ref{fig:mabo.vel007}
shows the velocity distribution
of the compiled galaxy sample and the Maxwell-Boltzmann distribution.
The distribution was calculated using galaxy velocities normalised
to the circular velocities, $v_{500}$, of their host clusters. The
Maxwell-Boltzmann distribution shown in the figure is given by
\be
\label{equ:mabo}
f(v) = N \sqrt{{2\over\pi}} \left({3\over\sigma^2_{\rm 3D}}\right)^{3/2} v^2 \exp \left(-{3v^2\over2\sigma^2_{\rm 3D}}\right)\ ,
\ee
where $\sigma_{\rm 3D}$ is the three dimensional velocity
dispersion. Here $\sigma_{\rm 3D}$ is calculated using normalised
galaxy velocities, $v/v_{500}$.  Note that the dashed line in the
figure is not a fit: it is the distribution given above calculated
with the galaxy velocity dispersion measured for the galaxies in our
compiled sample.
\begin{figure}
\epsfig{file=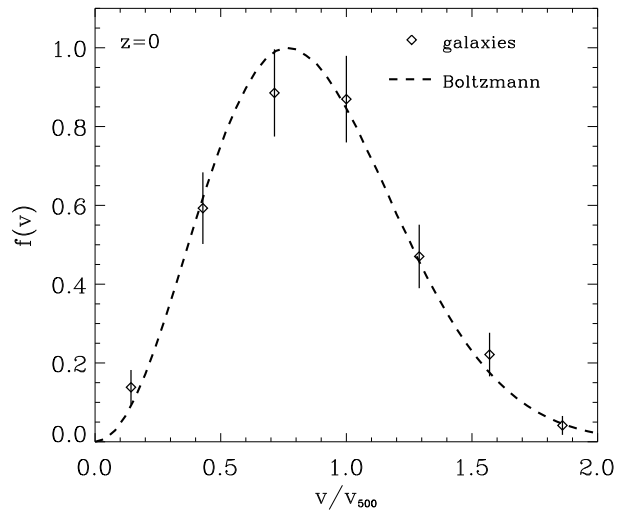,width=0.95\hsize}  
\caption{\label{fig:mabo.vel007} The normalised velocity distribution
  of the compiled galaxy sample ({\it diamonds}) and the
  Maxwell-Boltzmann velocity distribution ({\it dashed} line for all
  galaxies within $r<5r_{500}$). The error bars for galaxy
  distribution indicate the $1\sigma$ confidence interval.  Note that
  the Maxwell-Boltzmann distribution is not a fit. It is calculated
  using the measured velocity dispersion of the galaxy sample.}
\end{figure}

The figure shows that the velocity distribution of the galaxy sample
is well described by the Maxwell-Boltzmann distribution. One can
therefore derive the mean velocity of the galaxies from their three
dimensional velocity dispersion. We will use this relation below to
compute the mean velocity of galaxy motions.

\section{Galaxy Motions}
\label{sec:motions}
In this section, we analyse the motions of galaxies in the 
simulated clusters.  The analysis consists of three parts.
First, we measure the mean radial velocity, $v_r$, the three
dimensional velocity dispersion, $\sigma_{\rm 3D}$, and its radial and
tangential components, $\sigma_{r}$ and $\sigma_{t}$, of dark matter,
gas and galaxies directly from the simulations. The errors for the
galaxy measurements are computed using the $\sigma^2$ as the basic
values, for the 3D distribution as well as for the radial and
tangential components. Then the 67\% confidence interval is computed
and by means of error regression transformed to the actually shown
error range of the various $\sigma$ distributions (compare to
\citealt{colin_etal00}). Using the radial
profiles of $\sigma_{r}$ and $\sigma_{t}$ we can then compute the
radial dependence of the velocity anisotropy parameter, $\beta$ (see
eq.~[\ref{equ:beta}]).  We also measure the sound speed $c_s$ of gas
directly from the simulation.  Second, we perform analyses using the
equilibrium equations to model different components of galaxy
clusters: the Jeans equation for galaxies and the hydrostatic
equilibrium equation for the intracluster gas.  Using the above
measurements as inputs, we use the Jeans equations to solve for the
radial velocity dispersion profile of galaxies.  Similarly, we
apply the hydrostatic equilibrium equation to solve for the sound
speed of gas.  These results are then compared to the direct
measurements from the simulations to assess the validity of the
equilibrium equations in modelling kinematics of galaxies and gas.
Third, we study the average Mach number of galaxy motions as a
function of cluster-centric radius.  As we show below, we find that
galaxies on average move supersonically with the average Mach number
of $1.3-1.4$ throughout the cluster volume.
\subsection{Velocity dispersion, anisotropy, and sound speed profiles}

We measure the mean radial velocity, $v_r$, and the three dimensional
velocity dispersion profile, $\sigma_{\rm 3D}(r)$ (and its radial and
tangential components, $\sigma_{r}(r)$ and $\sigma_{t}(r)$) for dark
matter, gas and galaxies in radial bins centered on the minimum of
cluster potential after subtracting the peculiar velocity of the
cluster. The peculiar velocity of the cluster is calculated as 
the mass-weighted bulk velocity of dark matter enclosed within
$r_{500}$.  We scale these measurements by the circular velocity
$v_{500}$ measured at $r_{500}$ of each cluster.  After re-scaling,
we compute the average profiles for the sample of eight simulated
clusters at $z=0$.

The upper four panels of figure~\ref{fig:today.lin005} show the
profiles measured in the simulations. The net radial velocity is small
throughout the cluster for both the dark matter and gas, except in the
cluster core where there is a small net inward motion.  Galaxies, on
the other hand, show some net radial velocity in some radial bins.
However, the magnitude of the velocity, $\sim 0.2 v_{500}$, is small
compared to the velocity dispersion of the galaxies.  Note that we
find considerably smaller net radial velocity if we analyse each
cluster at an epoch near $z=0$, when it appears to be most relaxed.
This indicates that the net radial velocity seen in the $z=0$ sample
is likely due to incompletely erased galaxy groups within the
clusters.

We find that the three dimensional velocity dispersion of galaxies is
biased with respect to that of dark matter.  In other words, galaxies,
on average, move faster than the average speed of dark matter
particles in clusters. The velocity bias, $b_v=\sigma_{\rm
  gal}/\sigma_{\rm dm}$, is $b_v\approx 1.1$ within the virial radius
and disappears outside the virial radius of the cluster. Note that the
velocity bias we find for the stellar systems is smaller than the
velocity bias found for subhalos in dissipationless simulations,
$b_v\sim 1.2-1.4$
\citep{colin_etal00,ghigna_etal00,diemand_etal04,gao_etal04b}.
Interestingly, we find that the velocity bias comes entirely from the
tangential component of the velocity dispersion. As can be seen in
figure~\ref{fig:today.lin005}, the radial velocity dispersion of
galaxies and dark matter match well.

Figure~\ref{fig:today.lin005} also shows that gas within the virial
radius is not at rest but has non-zero velocity dispersion.  We find
that the 3D velocity dispersion of gas is approximately constant at
$\sigma_{\rm gas}\approx 0.4-0.5 v_{500}$.  In the inner regions of
clusters gas moves with typical velocity of $\sigma_{\rm gas}/\sqrt{3}
\sim 0.2-0.3 c_s$, but gas velocity dispersion becomes $\sim 0.5c_s$
outside the virial radius reflecting the increasing strength of the
infall motions and lesser degree of relaxation \citep*[see
also][]{nagai_etal03,sunyaev_etal03}.  Note that these random velocities of gas
contribute to the pressure support within clusters, in addition to the
support from thermal pressure.

The {\it bottom-left} panel of the figure~\ref{fig:today.lin005} shows
the radial profile of the velocity anisotropy parameter,
\be
\label{equ:beta}
\beta(r) = 1 - {\sigma^2_{t}(r)\over2 \sigma^2_{r}(r)}
\ee
(see, e.g., \citealt{binney:galactic}).  The figure shows that both
the dark matter and gas tend to have a slight radial anisotropy. The
anisotropy is nearly constant in magnitude: $\beta \approx 0.2$ for
dark matter (cp. \citealt{hoeft_etal04}) and $\beta \approx 0.4$ for
gas.  Note, however, the  velocity anisotropy of gas drops within
$\beta \approx 0.2$ in the centre.  For galaxies, we find similar
values of $\beta$ on the outskirts of clusters.  However, the value of
$\beta$ decreases gradually and reaches zero at $r\lesssim
0.5r_{500}$.  A fit to the actual measurements in simulations using
the fitting formula of \citet[][]{colin_etal00},
\be
\label{equ:colin}
\beta = \beta_m {4r\over r^2 + 4} + \beta_0\ ,
\ee
gives best fit parameters of  $(\beta_m,\beta_0)=(0.76,-0.39)$.
This fit is shown as a dashed line in figure~\ref{fig:today.lin005}. 
It describes the values measured in simulations quite well. 
Our best fit value of
$\beta_0$ is systematically smaller than the value found by
\cite{colin_etal00} in dissipationless simulations. This indicates
that velocities of galaxies in our simulations have more enhanced
tangential velocity dispersion compared to the subhalos in
dissipationless simulations.

The {\it bottom-right} panel in the figure~\ref{fig:today.lin005}
shows the average sound speed of intracluster gas.  We compute this
quantity in two ways.  First, we measure the average sound speed of
gas in the vicinity of galaxies, as sketched in the
figure~\ref{fig:measure}, and average over the measurements of all
galaxies in each radial bin.  Alternatively, we measure the average
sound speed of gas by simply computing the average of the sound speeds
measured in individual mesh cells in each bin.  The figure shows that
the average sound speeds calculated in these two ways agree well. Note
that the sound speed of gas is not constant but increases
monotonically toward the cluster centre.

\begin{figure}
\epsfig{file=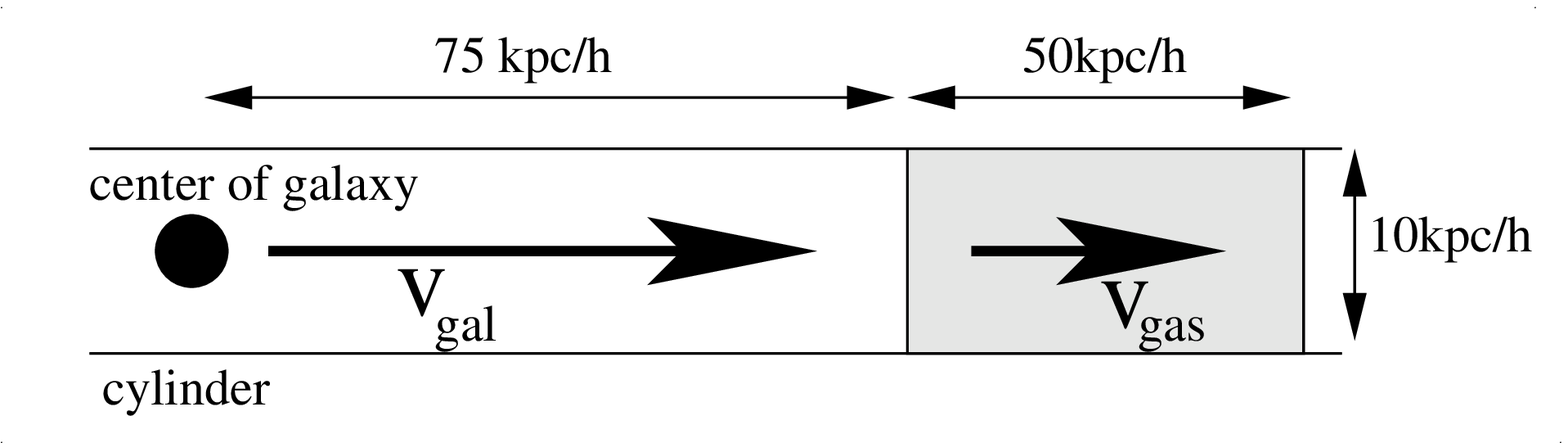,width=\hsize} 
\caption{\label{fig:measure} The sketch illustrates the measurements
of the sound speed of gas in the vicinity of galaxies.  We first
identify the centre of the galaxy (marked as the {\it solid circle}) and draw a
cylinder with the diameter of $10 h^{-1}{\rm kpc}$ aligned parallel to
the direction of galaxy motion.  We then measure the
average sound speed of gas in a segment of the cylinder with the size
of $50 h^{-1}{\rm kpc}$ located $75 h^{-1}{\rm kpc}$ ahead of the
galaxy centre.  To estimate the Mach number for a galaxy, we use the
relative velocity of the galaxy centre and the intracluster gas, ${\rm
v_{gal}-v_{gas}}$, as a measure of the galaxy speed. }
\end{figure}
\begin{figure*}
\begin{center}
\epsfig{file=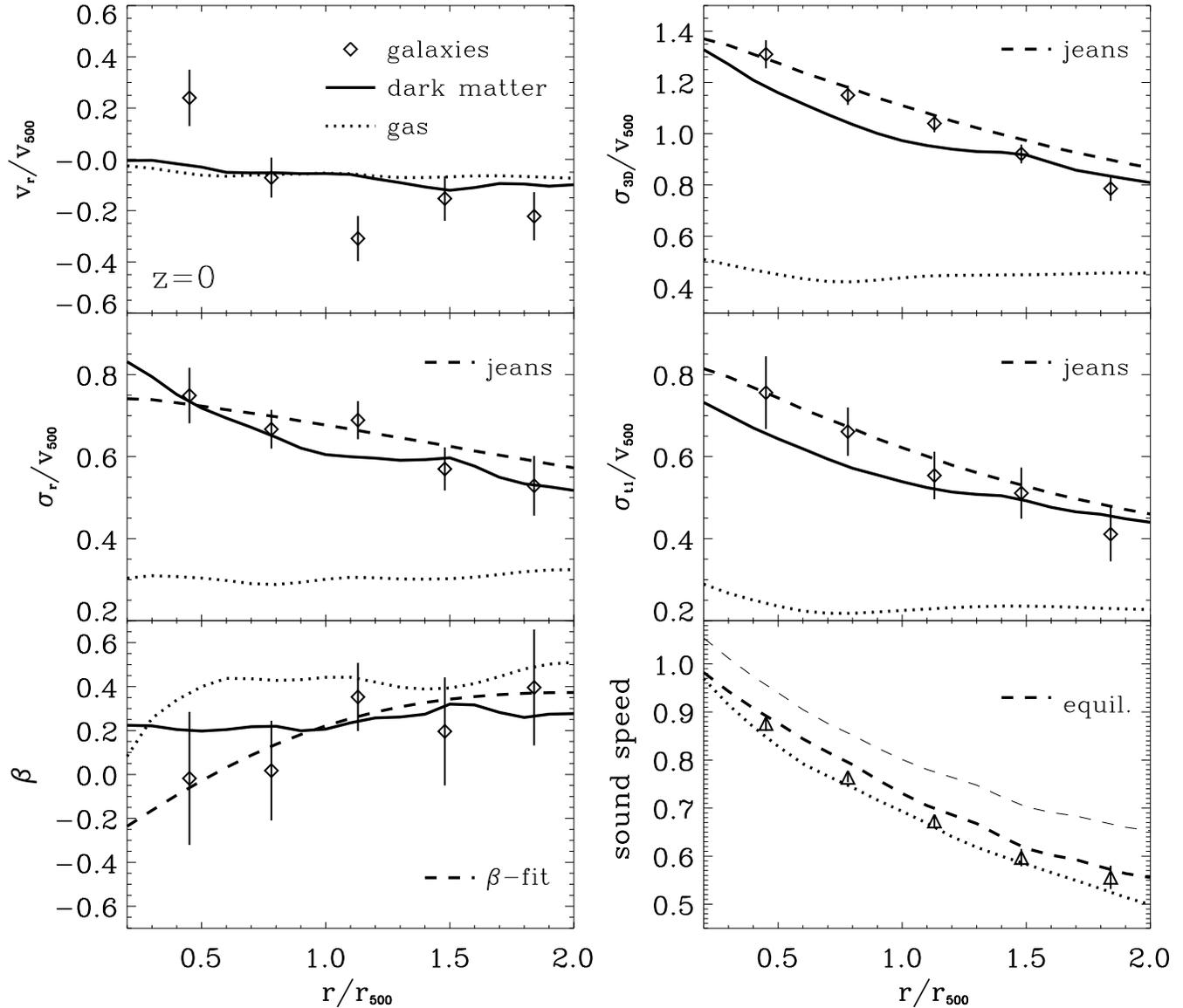,width=\hsize} 
\end{center}                    
\caption{\label{fig:today.lin005} The radial profiles of kinematic
  properties of dark matter ({\it solid}), gas ({\it dotted}) and
  galaxies ({\it diamonds}), averaged over the sample of eight
  simulated clusters at $z=0$.  Errorbars show 1 $\sigma$ errors of
  the mean. {\it Upper-four panels:} the mean radial velocity $v_{r}$
  ({\it top-left}), the three dimensional velocity dispersion
  $\sigma_{\rm 3D}$ ({\it top-right}), and its radial and tangential
  components, $\sigma_{r}$ ({\it middle-left}) and $\sigma_{t}$ ({\it
  middle-right}).  The {\it dashed} lines in these panels are the
  results derived using the Jeans equation.  {\it Bottom-left:} the
  anisotropy parameter $\beta$.  The {\it dashed} line is the best-fit
  $\beta$ profile of galaxies based on the
  formula~(\ref{equ:colin}). {\it Bottom-right: } the average sound
  speed of intracluster gas $c_s$.  The {\it triangles} are the
  measurements of sound speed of gas ahead of galaxies, as sketched in
  Fig.~\ref{fig:measure}. The {\it dotted} line is an average of the
  sound speeds in individual mesh cells. The two dashed lines show the
  solutions of the hydrostatic equilibrium equation, where we assumed
  $\sigma_{\rm turb}=0$ ({\it thin dashed} line) and $\sigma_{\rm
  turb}=\sigma_{\rm 3D}$[gas] ({\it thick dashed} line). }
\end{figure*}

\subsection{The equilibrium analysis}
In this section, we apply the two equilibrium equations, the Jeans
equation for galaxies and the hydrostatic equilibrium equation for
intracluster gas to calculate velocity dispersion profiles under the
assumptions that clusters are spherically symmetric and are in
equilibrium.  We then compare the results to the direct measurements
of the velocity dispersion from the simulations to assess the validity
of the equilibrium assumption.

\subsubsection{The Jeans equation}
\label{sec:jeans}
We apply the Jeans equation to derive the average velocity dispersion
of galaxies in galaxy clusters. For the relaxed systems with no
rotational support ($\bar{v}_r = \bar{v}_\theta = 0$), the
Jeans equation is
\be
\label{equ:jeans}
{1\over n_{\rm gal}}{\frac{d}{{d}r}}\left(n_{\rm gal}\sigma^2_r\right)
+
2\beta{\sigma^2_r\over r} 
= 
-{{d}\Phi\over{d}r}
\ee
where $\beta$ is the velocity anisotropy parameter in
Eq.~(\ref{equ:beta}), $n_{\rm gal}$ denotes the number density of
galaxies and $\Phi(r)={\rm G}M_{\rm tot}(r)/r$ is the gravitational
potential of the cluster. Integration of the Jeans equation requires
knowledge of (1) the total mass profile, $M_{\rm tot}$, (2) the number
density profile of galaxies, $n_{\rm gal}$ and (3) the radial profile
of the velocity anisotropy, $\beta(r)$.  In order to solve the
equation for the sample of clusters using the average profiles, it is
convenient to work with the dimensionless variables.  We, therefore,
scale each variable using their values at $r_{500}$ and then construct
an average profile for all eight clusters.  For the number density of
galaxies, we use the best-fit NFW profile discussed in
\S~\ref{sec:nrden} (see also Fig.~\ref{fig:dens.lin010}).  For the
velocity anisotropy profile, we use the fit to the actual measurements
in simulations using the fitting formula of \citet[][see
eq.~(\ref{equ:colin}) above]{colin_etal00}.

Using the derived values of $\sigma_r (r)$ and the best-fit $\beta(r)$
profile, we compute the tangential and total velocity dispersion
profiles, $\sigma_t(r)$ and $\sigma_{\rm 3D}(r)$. These profiles are
shown by the {\it dashed} lines in the corresponding panels of
figure~\ref{fig:today.lin005}.  All equilibrium solutions agree well
with the values directly measured in simulations.

\subsubsection{Hydrostatic equation}
For a system in equilibrium, the hydrostatic equation
\be
\label{equ:equil}
{1\over\rho_{\rm gas}}{dp_{\rm gas}\over dr} = - {d\Phi\over dr} \ 
\ee
relates the gas density
$\rho_{\rm gas}$ and gas pressure $p_{\rm gas}$, to the gravitational
potential of the cluster, $\Phi(r)$.  The gas sound speed is given
by
\begin{equation}  
\label{equ:sound}
c_s = \sqrt{\gamma
\left({p_{gas}\over\rho}-{\sigma^2_{\rm turb}\over3}\right)}\ .
\end{equation}
Note that this expression takes into account the additional
contribution to the pressure due to turbulent gas motions, where
$\sigma_{\rm turb}$ denotes the three dimensional velocity dispersion
of turbulent gas motions.

Using the hydrodynamical variables and the total mass profile $M_{\rm
tot}(r)$ measured directly from the simulations, we solve for the
sound speed of gas using the Eq.~(\ref{equ:equil}) and
(\ref{equ:sound}).  To solve these equations, we perform similar
scaling and interpolation discussed in the previous section on the
measured quantities.  The {\it dashed} lines in the {\it bottom-right}
panel of the figure~\ref{fig:today.lin005} show the results of the
calculation. The {\it thin} and {\it thick dashed} lines show the
sound speed calculated assuming $\sigma_{\rm turb}=0$ and $\sigma_{\rm
turb}=\sigma_{\rm 3D}$[gas] respectively, where $\sigma_{\rm 3D}$[gas]
is the velocity dispersion of the gas shown by the dotted line in the
top-right panel.  The figure shows that if random motions of gas are
ignored, the sound speed is overestimated by $\approx 10\%$.  This
indicates that random motions contribute $\gtrsim 10\%$ to the total
pressure support in simulated clusters, which is in agreement with
recent observations \citep[][]{schuecker_etal04}.  If we assume
$\sigma_{\rm turb}=\sigma_{\rm 3D}$, the estimated sound speed of gas
is in good agreement with the values measured directly from the gas
density and thermal pressure.
\subsection{The average Mach number of galaxy motions}
\label{sec:mach}
To estimate the Mach number of galaxies, $M \equiv \bar{v}/c_s$ , we
need to measure the average velocity of galaxy motions, $\bar{v}$, and
the average sound speed of the intracluster gas, $c_s$.  Below, we
compute the Mach number of galaxies in two different ways.  First, we
measure the average Mach number of galaxies directly from the
simulation by measuring the velocity of each galaxy and sound speed of
gas immediately ahead of the galaxy in the direction of its motion, as
sketched in figure~\ref{fig:measure}.  Once the Mach number is
measured for all galaxies, we compute the average Mach number in the
spherical bins and average over all clusters in the sample.

Alternatively, we can measure the average Mach number of galaxies
using the results of the Jeans and hydrostatic equilibrium equations
in the previous section.  As we discussed in \S~\ref{sec:mabo} 
(see Fig.~\ref{fig:mabo.vel007}), the velocity distribution of 
galaxies is well-described by the 
Maxwell-Boltzmann distribution.  This means that the mean velocity of
galaxies can be calculated using their three dimensional velocity
dispersion as
\begin{equation}
\bar{v} = 0.92 \sigma_{\rm 3D}.
\label{equ:vel1}
\end{equation}
We thus compute $\bar{v}$ from $\sigma_{\rm 3D}$ obtained by solving
the Jeans equation.  Similarly, we use the sound speed of gas, $c_s$,
obtained from the hydrostatic equation.  Note that estimating the
sound speed of gas requires the knowledge of turbulent gas motions
$\sigma_{\rm turb}$.  Since characterising the nature of the turbulent
gas motions is beyond the scope of the paper, we simply assume
$\sigma_{\rm turb}=\sigma_{\rm 3D}$[gas], which has been shown to
result in good agreement between the calculated and measured values of
the sound speed in Fig.~\ref{fig:today.lin005}.

Figure~\ref{fig:mach.lin005} shows the average Mach number of galaxies
in the simulated clusters.  The diamonds are the direct measurement
from the simulation, and the dashed line is calculated using solutions
of the equilibrium equations.  The figure shows that on average
galaxies move supersonically with the average Mach number of $\sim
1.3-1.4$ throughout the cluster volume.

\begin{figure}
\begin{center}
\epsfig{file=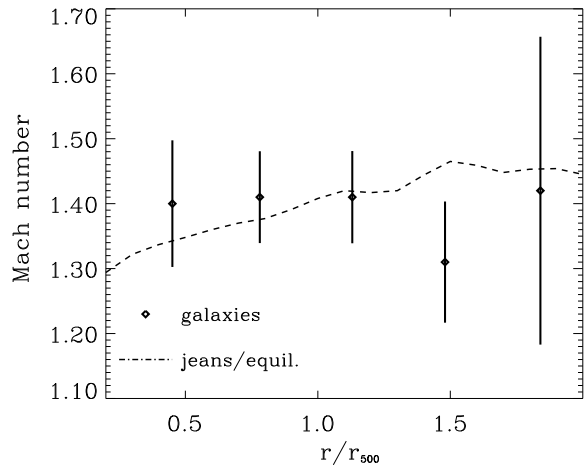,width=0.9\hsize}  
\end{center}                    
\caption{\label{fig:mach.lin005} The Mach number of galaxies orbiting
in the cluster environment. The {\it diamonds} show the direct
measurements from the simulations, averaged over the sample of eight
simulated clusters at $z=0$, and the error bars indicate the $1\sigma$
confidence interval.  The {\it dashed} line is the result of the
equilibrium equations, where we assumed that $\sigma_{\rm
turb}=\sigma_{\rm 3D}$[gas] to account for the turbulent gas motions
in the estimate of the sound speed of intracluster gas. The figure
shows that galaxies move supersonically with the Mach number of
$1.3-1.4$ throughout the cluster environment. }
\end{figure}

In the first approximation, the slightly supersonic motions of
galaxies can be explained by the following simple argument. The mean
velocity of galaxies is given by eq.~(\ref{equ:vel1}), while the sound
speed is $v_s=\left(\gamma P_{\rm gas}/\rho\right)^{1/2}$
$=\left(\gamma \sigma_{\rm gas}^2/3\right)^{1/2}$, where $\sigma_{\rm gas}^2$
is the velocity dispersion of the microscopic thermal motions of gas
particles.  If the three-dimensional velocity dispersions of gas particles and
galaxies are approximately equal, as is expected in equilibrium, we have:
\begin{equation}
\langle M\rangle=\bar{v}/c_s\approx 0.92\sqrt{3/\gamma}\approx 1.23,
\label{eq:machgamma}
\end{equation}
for $\gamma=5/3$. As was shown in the previous section, turbulent
pressure contributes $> 10\%$ to the total pressure and the sound
speed is actually $\approx 10\%$ lower than what it would be without
the turbulence. This can explain the average Mach number that we
measure $\langle M\rangle \approx 1.23/0.9\approx 1.37$.

\section{Discussion and Conclusions}
\label{sec:discussion}

We presented analysis of the galaxy motions in clusters using
high-resolution cosmological simulations of the concordance flat
$\Lambda$CDM model.  The simulations follow dynamics of dark matter
and gas and include various physical processes critical for galaxy
formation: gas cooling, heating and star formation.  These
simulations, therefore, follow the formation of galaxies and their
evolution in the dense cluster environment in a realistic cosmological
context.  Analysing motions of galaxies and the properties of
intracluster gas in the sample of eight simulated clusters at $z=0$,
we study velocity dispersion profiles of the dark matter, gas, and
galaxies. We measure the mean velocity of galaxy motions and gas sound
speed as a function of radius and calculate the average Mach number of
galaxy motions.

Our simulations show that galaxies, on average, move supersonically
throughout the cluster volume.  The average Mach number is $\langle
M\rangle\approx 1.4$, approximately independent of cluster-centric
radius.  The value of $\langle M\rangle$ can be attributed to the
difference between the three-dimensional velocity of galaxies and
one-dimensional sound speed and to the existence of the turbulent
motions of gas (see eq.~\ref{eq:machgamma}). The thermal pressure and,
hence, the sound speed are smaller than would be required if all of
the gas pressure support was due to thermal pressure. Also, gas and
galaxies have somewhat different radial density and velocity
anisotropy profiles.  The motions of galaxies and gas particles are
governed by the same potential, but non-zero velocity anisotropy and
differences in the radial distribution can lead to different velocity
dispersion and sound speed profiles.

We also find that galaxies move slightly faster than the dark matter
particles in clusters, although the magnitude of the velocity bias,
$b_v\approx 1.1$, is considerably smaller than the bias estimated for
subhalos in dissipationless simulations
\citep{colin_etal00,ghigna_etal00,diemand_etal04,gao_etal04b}.
Interestingly, we find velocity bias in the tangential component of
the velocity dispersion, but not in the radial component.  Despite the
small sample size used in this analysis, we find that these results
are robust and statistically significant.  Namely, the results do not
change if we analyse clusters by removing two of the least relaxed
clusters at $z=0$ or using clusters at the epoch when they appear most
relaxed.  Nevertheless, it would be good to verify these results with
higher resolution simulations and better statistics.

Our simulations show that the difference between the dark matter and
galaxy velocity dispersions is significant only in the inner regions
of clusters ($r/r_{500}\lesssim 0.7$).  This is also where
we find a corresponding difference in the velocity anisotropies.  With
these results in mind, we conjecture that the bias is likely caused by
the preferential disruption of objects on highly radial orbits.  In
other words, galaxies on radial orbits are prone to more efficient
tidal disruption than those on circular orbits with the same orbital
energy, since galaxies on radial orbits pass closer to the dense
central regions during their peri-centric passages.  If the effect is
significant, we expect the fraction of galaxies with the small (large)
tangential (radial) velocity component to decrease in the inner
regions of clusters.  The distributions of $v_r$ ($v_t$), therefore,
become progressively skewed towards smaller (larger) values with
decreasing radius.  

Our simulations show that this is indeed the case.
Specifically, we find that the distributions of $v_r$ and $v_t$
become increasingly non-gaussian, and $v_t/v_r$ ratio increases
monotonically at the smaller cluster-centric distances within
$r<r_{500}$.  Furthermore, if galaxies accrete onto clusters with the
positive orbital velocity anisotropy, $\beta(r)>0$ (i.e., preference
for the radial orbits), similar to dark matter particles, the
tidal disruption would also drive the system toward more isotropic
orbits. We indeed find that galaxy motions in the central regions of
the clusters are nearly isotropic ($\beta\approx 0-0.1$), while the DM
particles have the radial anisotropy of $\beta\approx 0.2-0.3$ (see
Figure~\ref{fig:today.lin005}).  

We also find considerable random bulk motions of gas.  The 3D velocity
dispersion is approximately constant as a function of radius:
$\sigma_{\rm gas}\approx 0.4-0.5 v_{500}$. In terms of the sound
speed, the gas moves with the typical velocity of $\sigma_{\rm
  gas}/\sqrt{3} \sim 0.2-0.3 c_s$ in the inner regions of clusters
\citep[see also][]{nagai_etal03,sunyaev_etal03}. Outside $r_{500}$, the typical
velocities are $\sim 0.5c_s$, which reflects the increasing strength
of the infall motions and lesser degree of relaxation.  The random
motions of gas contribute to the pressure support of galaxy clusters,
in addition to the support from thermal pressure.  We show that in our
simulations, random motions contribute $\approx 10\%$ of the total
pressure support. Recent X-ray observations of the Coma cluster also
show evidence of random motions of gas of a similar magnitude
\citep{schuecker_etal04}. The non-thermal pressure support, if neglected,
will bias measurement of the total mass in the hydrostatic 
analyses of the X-ray cluster observations by $\sim 10-20\%$. 

The supersonic motions of cluster galaxies may be an important source
of heating of the intracluster gas. Supersonically moving galaxies and
groups can drive weak bow shocks and shock-heat the ICM
\citep[e.g.][]{markevitch_etal04,finoguenov_etal04} or deposit energy via
viscous dissipation of soundwaves \citep{ruszkowski_etal04}. It is likely,
however, that a more efficient heating mechanism is the transfer of
the orbital energy of galaxy motions to the internal energy of gas via
dynamical friction.  Dynamical friction in the gaseous medium is
efficient only if the perturber moves supersonically
\citep{ruderman_spiegel71,rephaeli_salpeter80,just_etal90,ostriker99}.
Our results therefore confirm that this is a potentially viable
heating mechanism.  \cite{elzant_etal04} have recently argued that
the dynamical friction heating is sufficiently efficient to prevent
formation of the cooling flows in cluster cores. Their estimates show
that in the Perseus cluster galaxies do move supersonically with the
average Mach number of $M\sim 1.5-2$. For reasonable choices of the
galaxy mass-to-light ratios the rate of energy loss to dynamical
friction is sufficient to offset a cooling of gas in this cluster.
\cite{elzant_etal04} also emphasise that this mechanism is self-regulating:
as gas is heated, the galaxy motions become subsonic and heating
becomes inefficient.

In principle, dynamical friction heating should be modelled
self-consistently in cosmological simulations. Our simulations,
however, do suffer from the well-known ``overcooling problem:'' the
fraction of baryons in stars and cold gas is at least a factor of two
higher than observed for the systems of the mass range we consider.
For example, the typical star formation rate in the cool cores in the
central galaxy in our clusters is $\gtrsim 1000\ {\rm M}_{\odot} {\rm
  yr}^{-1}$, which is much larger than is allowed by the observed
values of the mass accretion rates \citep[see e.g.,][]{kaastra_etal04,
  peterson_etal01, peterson_etal03, tamura_etal01}. This 
indicates that the dynamical friction heating is not efficient in our
simulations. This could be due to their limited spatial resolution (a
few kpc). It is possible, that to resolve the dynamical friction wakes
properly a sub-kiloparsec resolution is needed. However, it is also
possible that the wake formation is prevented by the random motions of
the gas that we find in our simulations. These motions are much
stronger than the gravitational pull of any individual galaxy and it
is likely that galaxies simply cannot form wakes in such highly
chaotic velocity field.

Given that dynamical friction can provide an attractive source of the
ICM heating, it will be important to pursue this subject further. The
progress can come both from the higher resolution cosmological
simulations and from controlled gasdynamics experiments of a
gravitating body moving in a gas flow. It would be interesting, for
example, to test whether dynamical friction is equally efficient for
a perturber moving in laminar and strongly turbulent flows.

\section*{Acknowledgements}
We are grateful to Avi Loeb for pointing out the simple explanation
for the $M>1$ measurement to us. We would also like to thank the
anonymous referee for helpful and constructive comments. This work was
supported by the National Science Foundation (NSF) under grants No.
AST-0206216 and AST-0239759, by NASA through grant NAG5-13274, and by
the Kavli Insitute for Cosmological Physics at the University of
Chicago. D.N.  is supported by the NASA Graduate Student Researchers
Program and by NASA LTSA grant NAG5--7986. We would like to thank
NSF/DAAD for supporting our collaboration. AVK would like to thank
Aspen Center for Physics and organisers of the ``Starformation in
galaxies'' workshop for hospitality and productive atmosphere during
completion of this paper.  The cosmological simulations used in this
study were performed on the IBM RS/6000 SP4 system at the National
Center for Supercomputing Applications (NCSA) and at the Leibniz
Rechenzentrum Munich and the John von Neumann Institute for Computing
J\"ulich.

\bibliography{literature}
\end{document}